\documentstyle[12pt]{article}
\begin{document}
hep-th/9610052\\
\par
\begin{center}
{\bf BRANCHED POLYMERS WITH LOOPS}\\
\vspace{0.8cm}
J. Jurkiewicz\footnote{Permanent address: Institute of Physics,
 Jagellonian University, ul. Reymonta 4, 30-059 Krak\'ow, Poland.} 
and A. Krzywicki\\
\vspace{0.8cm}
LPTHE, B\^{a}t. 211, Universit\'e Paris-Sud, 91405 Orsay, 
France\footnote{Laboratoire associ\'e au CNRS}\\
\vspace{1.3cm}
{\bf Abstract}
\end{center}
We propose a classification of critical behaviours of
branched polymers for arbitrary topology. We show that
in an appropriately defined double scaling limit the
singular part of the partition function is universal. 
We calculate this partition function exactly in the generic
case and perturbatively otherwise. In the discussion 
section we comment on the relation between branched
polymer theory and Euclidean quantum gravity.
\vfill
\par\noindent
October 1996\\
LPTHE Orsay 96/82
\pagebreak
\section{Introduction}
The statistical mechanics of branched 
polymers (BP) is one of the 
simplest and most tractable models 
of random geometry. It is a
subject of intrinsic interest and has 
already been studied by several
authors \cite{bp1}-\cite{bp2}. A further 
motivation for developing 
these studies is provided by the 
recent suggestion \cite{bbpt} that important
features of Euclidean quantum gravity 
can be inferred from those of 
the ensemble of branched polymers 
isomorphic to the ensemble of trees 
of baby universes connected by wormholes. 
\par
The main body of this paper is 
devoted to the discussion of the
critical properties of naked\footnote{That
is without matter fields living on them.} BP models, 
including the possibility of
loop formation. Some results have 
been obtained earlier by other
authors and are included here for 
completeness. We shall comment 
about quantum gravity in the discussion section.
\par
We formulate the problem as a minifield 
theory, defining a generating
function $W(j)$ by the familiar equation:
\begin{equation}
e^{\Lambda W(j)} = \int d\phi e^{\Lambda [
 - {\lambda \over 2} \phi^2
+ V(\phi) +j \phi]}
\label{defw}
\end{equation}
\noindent
However, here the integration variable 
$\phi$ is just a real
number. The interaction potential 
is assumed to have the form 
\begin{equation}
V(\phi) = \sum_k (p_k / k) \phi^k \; , \; k > 2
\label{defv}
\end{equation}
\noindent
with positive couplings: $p_k \ge 0$. 
Thus, strictly speaking, the
integral in (\ref{defw}) does not exist. 
It is introduced in the
first place to define a perturbation series. 
Obviously, one can associate diagrams with 
the terms of this series. The propagator
corresponding to a link is $\lambda^{-1}$, 
a source $j$ is attached 
to each external link and the number of 
loops in the diagram is identical to the 
power of $\Lambda^{-1}$. The generating function
of {\em rooted} diagrams (those with one 
marked external point) is
given by $\partial W /\partial j$.

\section{The trees}
The saddle point $\phi = Z$ is 
found from the equation
\begin {equation}
Z = \lambda^{-1} [ j + \sum_k p_k Z^{k-1}]
\label{saddle}
\end{equation}
\noindent
When calculated in the saddle point 
approximation, the functions $W = W_0$  
and $\partial W_0 /\partial j$ generate the tree and 
the rooted tree diagrams respectively. Since
\begin{equation}
Z = \partial W_0 /\partial j \; ,
\label{rooted}
\end{equation}
\noindent
it is evident that $Z$ is the partition 
function of rooted BP. And indeed 
eq. (\ref{saddle}) is generally used 
as the defining equation for rooted 
BP. We write here $j$ instead of $p_1$ 
and we have, for definiteness, 
incorporated $p_2$ into $\lambda$. 
But this is just a matter of conventions.
\par
Introduce the positive definite function
\begin{equation}
F(\phi) = {j \over \phi} + \sum_k p_k \phi^{k-2} \; 
\label{deff}
\end{equation}
\noindent
and rewrite (\ref{saddle}) as
\begin{equation}
\lambda = F(Z)
\label{saddle2}
\end{equation}
\noindent
The positivity of $p_k$ implies that 
$F'(\phi$) can have at most
one zero for $\phi > 0$ and that 
$F''(\phi) > 0$. Let $r$ denote
the radius of convergence\footnote{Possibly infinite,
but not infinitesimally small.} 
of the series in (\ref{deff}) and let
$\phi = \bar{\phi}$ be the point in the 
interval $(0,r]$ where
$F(\phi)$ takes its minimum value. 
We deduce from (\ref{saddle2})
that the parameter $\lambda$ cannot 
decrease below $\lambda_c = F(\bar{\phi})$. 
Consequently, the partition function
$Z$ must have a singularity at 
$\lambda = \lambda_c$.
\par
Write $\delta \lambda = \lambda - \lambda_c$ 
for later convenience and assume that in the 
neighbourhood of $\delta \lambda =0$ the
singular part of the partition function behaves as
\begin{equation}
Z_{\mbox{\rm s}} \sim 
\delta \lambda^{1 - \gamma} \; ,
\label{susc}
\end{equation}
\noindent
This is the conventional definition 
of the (geometrical) susceptibility exponent
$\gamma$. Another interesting quantity is the
two-point correlation function $C(x)$,
where $x$ is the (integer) distance between two
marked points. It has been shown in \cite{adj} 
that at the tree level this function is
\begin{equation}
C(x) \sim [\lambda^{-1} {{\partial} \over {\partial Z}}(ZF(Z)]^x 
\label{corr}
\end{equation}
For large $x$
\begin{equation}
x^{-1} \log{C(x)} = - \; \mbox{\rm const} \;
\delta \lambda^{1 \over {d_H}} 
\label{corr2}
\end{equation}
\noindent
where $d_H$ is the Hausdorff dimension.
\par
Generalizing the considerations 
of ref. \cite{bb1} one can present the 
catalogue of possible critical behaviours: 
\par\noindent
{\bf The generic case.}
\par
In the generic situation $\bar{\phi} < r$. 
One then has $F'(\bar{\phi}) = 0$ and
\begin{equation}
F(Z) = F(\bar{\phi}) + 
{1 \over 2} F''(\bar{\phi}) (Z - \bar{\phi})^2 + ... \; ,
\label{gen1}
\end{equation}
\noindent
which after inversion yields\footnote{We 
have written a minus sign in front
of the 2nd term on the right-hand side 
because we consider the branch
$Z \le \bar{\phi}$ as the physical one.}
\begin{equation}
Z \simeq \bar{\phi} - [2 / F''
(\bar{\phi})]^{ 1 \over 2}
 \; \delta \lambda^{ 1 \over 2}
\label{gen2}
\end{equation}
\noindent
Hence $\gamma = { 1 \over 2}$. One also
finds $d_H = 2$.
\par
The alternative to the generic case occurs 
when $\bar{\phi} = r < \infty$. It follows from
the positivity of $p_k$ that $\phi = r$ is
a singular point of $F(\phi)$. Since 
$F(\phi)$ decreases monotonically 
towards $F(r) \ge 0$ one expects the singularity
to be a branch point in all cases of
physical interest\footnote{The interested reader can
consult \cite{dien} and in particular the theorems 
by Leau, Le Roy and Lindel\"{o}f.}.
Hence (modulo logs)
\begin{equation}
F(Z) = \sum_0^n {{(-1)^k} \over {k!}} 
F^{[k]}(r) (r - Z)^k
+ c (r - Z)^{\beta - 1} + ... \; , \; n+1 < \beta < n + 2
\label{atyp}
\end{equation}
\noindent
Since $F$ is concave, one must have 
$\beta > 2$. Furthermore, from the relation between
$F$ and $V$ and from the positivity of all the
derivatives of $V$ one easily deduces that 
$c (-1)^n < 0$. There are two possibilities:
\par\noindent
{\bf The semi-generic case.} 
\par
It occurs when $\bar{\phi} = r$ and $F'(r) <  0$. 
Inverting (\ref{atyp}) one finds
\begin{equation}
Z_{\mbox{\rm s}} \simeq 
{c \over {[-F'(r)]^\beta}} \; 
\delta \lambda^{\beta - 1}
\label{semi}
\end{equation}
\noindent
and therefore $\gamma = 2 - \beta$. In this case
$d_H = \infty$.
\par\noindent
{\bf The marginal case.}
\par 
It occurs when $\bar{\phi} = r$ 
and $F'(r) = 0$. For $2 < \beta < 3$, 
inverting eq. (\ref{atyp}) one finds
\begin{equation}
Z_{\mbox{\rm s}} \simeq \mbox{\large(}
{{\delta \lambda} \over c}
\mbox{\large)}^{1 \over {\beta-1}}
\label{marg}
\end{equation}
\noindent
so that $\gamma = (\beta - 2)/(\beta - 1)$. The
Hausdorff dimension is $d_H = (\beta - 1)/(\beta - 2)$. 
When $\beta > 3$ the situation is analogous to the
generic one:  one has (\ref{gen2}) with $r$
in place of $\bar{\phi}$ 
and $\gamma = { 1 \over 2}$ while $d_H=2$.
\par
To summarize, in the non-generic cases 
there is a continuum 
of universality classes,
characterized by the exponent $\beta$ 
and by the vanishing (or non-vanishing)
of $F'(r)$. Each class corresponds to 
an infinity of different choices
of the couplings $p_k$. 

\section{The loop expansion}
\subsection{The generic case}
The loop expansion is obtained calculating 
corrections to the saddle-point approximation. 
This can be done either directly or
by using Dyson-Schwinger equations. We employ 
here the second method, which is more elegant 
and enables one to write rapidly the
BP equation, an analogue of the string 
equation one finds in the
double scaling limit of 2d gravity \cite{bk}.
\par
Set $U(\phi) = {\lambda \over 2} \phi^2 - V(\phi)$. 
The Dyson-Schwinger equation reads
\begin{equation}
[U'(\Lambda^{-1}{{\partial} \over  {\partial j}}) - j] \;  
e^{\Lambda W(j)} = 0
\label{ds1}
\end{equation}
\noindent
which can be rewritten as
\begin{equation}
U'(Z + \Lambda^{-1} {{\partial} \over 
{\partial j}}) \cdot 1 = j
\label{ds2}
\end{equation}
\noindent
Here $Z = \partial W/\partial j$ denotes 
the full partition function, with loop 
corrections included. Let $\phi_0$ be the point 
where $U''$ vanishes: $U''(\phi_0) = 0$. Define
\begin{equation}
\Delta = 2 [j - U'(\phi_0)]/U'''(\phi_0)
\label{delta}
\end{equation}
\noindent
and the double scaling limit: 
\begin{equation}
\Delta \to 0 \; , \; 
\Lambda \to \infty \; , \; 
t = - {1 \over 3}\;  U'''(\phi_0)\;
 \Lambda\; \Delta^{3 \over 2} = 
\mbox{\rm const} 
\label{limit}
\end{equation}
\noindent
Both $\Delta$ and $t$ are positive definite. 
Set $Z = \phi_0 - \Delta^{1  \over  2} 
\; \chi(t)$ in (\ref{ds2})
and expand. In the limit (\ref{limit}) 
one gets the BP equation
promised at the beginning of this section:
\begin{equation}
1 = \chi^2(t) + {1 \over {3t}} \chi(t) + \chi'(t)
\label{bpeq}
\end{equation}
\noindent
From the Riccati equation (\ref{bpeq}) one 
can obtain $\chi$ as a universal power series 
in $t^{-1}$:
\begin{equation}
\chi(t) = 1 - {1 \over {6 t}} - {5 \over {72 t^2}} + ...
\label{expa}
\end{equation}
\noindent
We have assumed that the first term is $+1$. This
is the physical choice (cf. the footnote preceding
eq. (\ref{gen2})), which also guarantees that the terms
corresponding to higher topologies give a positive 
contribution to the partition function. The behaviour 
of the coefficient $\chi_n$ of $t^{-n}$ , for $n \gg 1$, 
can be easily estimated using (\ref{bpeq}):
\begin{equation}
\chi_n \sim \mbox{\rm const} \;
 {{\Gamma(n)} \over {2^n}}
\label{an}
\end{equation}
\noindent
The constant above can be found numerically and
is approximately equal to $- 0.32$. As one could 
expect, the series (\ref{expa}) is not Borel summable.
The perturbative series does not determine 
$\chi(t)$ uniquely. The first singularity of 
the Borel transform occurs when its
 argument equals $2$. Thus the
leading non-perturbative contribution to 
$\chi(t)$ is proportional to
$e^{-2t}$.
\par
Let $f(t)$ be the primitive of $\chi(t)$: 
$f'(t) = \chi(t)$. Setting
$\Phi(t) = e^{f(t)}$ and changing the independent 
variable $t \to z =({3 \over 2} t)^{2 \over 3}$ 
one gets from (\ref{bpeq})
the Airy equation
\begin{equation}
\Phi'' = z \Phi
\label{airy}
\end{equation}
\noindent
Hence
\begin{equation}
\Phi(z) = a \; \mbox{\rm Ai}(z) + 
b \; \mbox{\rm Bi}(z)
\label{phi}
\end{equation}
\noindent
and
\begin{equation}
\chi(t) = \left({3 \over 2} t\right)^{- {1 \over 3}} \;
 {{\Phi'(z)} \over {\Phi(z)}}
 \; ,  \; z =\left({3 \over 2} t\right)^{2 \over 3}
\label{exact}
\end{equation}
\noindent
The function $\chi(t)$ depends on a single, but 
arbitrary parameter $a/b$, which measures the strength of the
nonperturbative contribution to the solution\footnote{
In the limit $b \to 0$ one obtains the unphysical
solution with $\chi(\infty)= -1$.}. In this respect 
the situation resembles that of string theory.
\par
It remains to find the relation between 
$\Delta$ and $\delta \lambda$.
The position of the point $\phi = \phi_0$ 
depends on $\lambda$, but not on
$j$. One easily finds, however, that as 
$\lambda$ approaches $\lambda_0$,
its j-dependent critical value, one has
$\phi_0 \to \bar{\phi} + \delta \lambda 
/ V'''(\bar{\phi})$, which is also j-dependent. 
One further finds from the definition
of $\Delta$ that
\begin{equation}
\Delta = {{2\bar{\phi}} \over 
{V'''(\bar{\phi})}}\;  \delta \lambda
\; [1 + \mbox{\rm O}(\delta \lambda^2)]
\label{deltas}
\end{equation}
\noindent
The behaviour of $\chi(t)$ at large $t$ 
and the above equation imply the following 
dependence of the susceptibility
exponent on the number of loops $L$:
\begin{equation}
\gamma_L = {1 \over 2} + {3 \over 2} L
\label{gamgen}
\end{equation}
\noindent
a result obtained by a different method in \cite{dur}.

\subsection{The strong coupling regime}
Set $V(\phi) = g \bar{V}(\phi)$ and assume 
that $g \gg 1$. This defines the strong coupling 
regime. It is obvious from (\ref{deff})
that in this regime $\bar{\phi} \sim \sqrt{j/g}$ 
and $V'''(\bar{\phi}) \sim g$. Hence, for large 
enough $g$ one necessarily has $\bar{\phi} < r$ 
and one is in the generic situation. One easily 
finds
\begin{equation}
t \sim {{j^{3 \over 4}} \over {g^{5 \over 4}}} \; 
 \Lambda \;  \delta \lambda^{3 \over 2}
\label{tstr}
\end{equation}
\noindent
Therefore, in the double scaling limit $t$ remains a
small parameter. Using (\ref{exact}) one gets
\begin{equation}
Z_{\mbox{\rm sing}} = - \sqrt{\Delta} \chi(t) 
\sim - {{j^{1 \over 4}} \over 
{g^{1 \over 3}}} \;  \sqrt{\delta \lambda} 
\label{zstr}
\end{equation}
\noindent
The free parameter $a/b$ is hidden in the 
coefficient multiplying the right-hand side 
of (\ref{zstr}). The susceptibility exponent 
$\gamma = {1 \over 2}$ and the number of 
loops is zero\footnote{The average number of 
loops equals $\partial \ln{Z}/\partial ln{N}$ 
calculated at fixed $\delta \lambda$.}.

\subsection{The non-generic cases}
When the minimum value of $\lambda$ is found at the 
boundary of the convergence interval of the potential,
the equation $U''(\phi_0)=0$ has no solution with
$\phi_0 \leq r$. In this case we combine the Dyson-Schwinger
equation(\ref{ds2}) with the expansion (\ref{atyp}). We then
encounter formal expressions of the type 
$[h(x) + \partial/\partial x]^{\beta -1}$, which must be given
a precise meaning. Write
\begin{equation}
[h(x) + \partial/\partial x]^{\beta -1} =
{1 \over {\Gamma(1-\beta)}} \int_0^{\infty} ds \; 
s^{-\beta} e^{- s [h(x) + \partial/\partial x]}
\label{integ}
\end{equation}
\noindent
Define $G$ by
\begin{equation}
e^{- s [h(x) + \partial/\partial x]} = 
e^G e^{- s \partial/\partial x}
\label{gdef}
\end{equation}
\noindent
It is easy to check that $e^G$ satisfies the
differential equation
\begin{equation}
{{\partial} \over {\partial s}} e^G = - e^G 
e^{s \partial/\partial x} h(x) 
e^{- s \partial/\partial x}\equiv - e^G h(x-s)
\label{difg}
\end{equation}
\noindent
with the boundary condition $G=1$ at $s=0$.
Hence $G$ is a c-number function:
\begin{equation}
G = H(x-s) - H(x)
\label{gsol}
\end{equation}
\noindent
where $H(x)$ is the primitive of 
$h(x)$: $H'(x) = h(x)$. Using this result we obtain
\begin{equation}
[h(x) + \partial/\partial x]^{\beta -1} \cdot 1 =
\sum_{k=0}^{\infty} {{\Gamma(1- \beta + k)}
\over {\Gamma(1- \beta)}} h^{\beta - 1 -k} R_k(h) \;
\label{ser}
\end{equation}
\noindent
where $R_k(h)$ is defined by
the equation 
\begin{equation}
R_k(h) = {{e^{- H(x)}} \over {k!}} {{\partial^k}
\over {\partial s^k}} \mbox{\large[} e^{H(x-s)}e^{sh(x)}
\mbox{\large]}_{s=0}
\label{rn}
\end{equation}
\noindent
and depends on the derivatives 
$h'(x), ... , h^{[n-1]}(x)$. Using Leibniz 
formula to calculate the derivative in (\ref{rn}) and 
replacing $\partial /\partial s$ by $\partial /\partial x$
in the appropriate place one can get rid of $s$. After some 
algebra one obtains
the following recursion formula
\begin{equation}
R_k(h) = {1 \over k} [ h'(x) R_{k-2}(h) - R'_{k-1}(h)]
\label{rec}
\end{equation}
\noindent
with $R_0 = 1$ and $R_1 = 0$. As one might expect 
the series (\ref{ser}) would be
truncated at $k = \beta$ if $\beta$ were an integer
(which it is not!). We shall now use the above results
to study the non-generic scaling behaviour.
\par\noindent
{\bf The semi-generic case}
\par
We define 
\begin{equation}
\Delta = U'(r) - j \equiv r \delta \lambda
\label{del2}
\end{equation}
\noindent
and define the scaling limit as follows:
\begin{equation}
\Delta \to 0 \; , \; 
\Lambda \to \infty \; , \; 
t = {{\Lambda \Delta^2}\over{-rF'(r)}} = \mbox{\rm const} 
\label{limit2}
\end{equation}
\noindent
Rewrite the Dyson-Schwinger eq. (\ref{ds2}) in terms
of $F$, set $Z = r - f$ and use the expansion (\ref{atyp})
to get
\begin{eqnarray}
0 =\Delta + r F'(r) f 
  - \;  r c (f + {{\partial} \over
 {\Lambda \partial \Delta}})^{\beta-1} \cdot 1 + ...
\label{eq1}
\end{eqnarray}
\noindent
where the dots represent terms irrelevant in the scaling limit.
\par
We have $f = f_{\mbox{\rm a}} - Z_{\mbox{\rm s}}$, where
\begin{equation}
f_{\mbox{\rm a}}= {{\Delta} \over {-rF'(r)}} + ... 
\label{fa}
\end{equation}
\noindent
is analytic in $\Delta$. In the case under consideration, 
$Z_{\mbox{\rm s}} \sim \Delta^{\beta -1}$ is subleading, 
compared to $f_{\mbox{\rm a}}$, because $\beta > 2$.
But we are precisely interested in this subleading term.
\par
In the scaling limit one gets from (\ref{eq1})
\begin{equation}
Z_{\mbox{\rm s}} \simeq {c \over {-F'(r)}} 
(f + {{\partial} \over
 {\Lambda \partial \Delta}})^{\beta-1} \cdot 1 
\label{eq2}
\end{equation}
\noindent
We can now use (\ref{ser}), slightly 
modified: because of the factor
$\Lambda^{-1}$ in front of ${{\partial} \over
 {\partial \Delta}}$ one has 
$\tilde{R}_k(f) = \Lambda^{-k} R_k(\Lambda f)$ 
instead of $R_k(f)$: 
\begin{equation}
Z_{\mbox{\rm s}} \simeq {c \over {-F'(r)}} \sum_{k=0}^{\infty}
{{\Gamma(1- \beta + k)}
\over {\Gamma(1- \beta)}} f^{\beta - 1 - k}
\Lambda^{-k} R_k(\Lambda f)
\label{eq3}
\end{equation}
\noindent
One finds by inspection that in the scaling limit only
the terms with even $k$ contribute to the right-hand side.
The leading contribution comes from the first term in
\begin{equation}
R_{2m} = c_{2m} (f')^m + ...
\label{eq4}
\end{equation}
\noindent
We find from the recursion equation (\ref{rec}) that 
\begin{equation}
c_{2m} = {1 \over {2^m m!}}
\label{eq5}
\end{equation}
\noindent  
Using (\ref{fa}), (\ref{eq4}) and (\ref{eq5}) we 
finally get\footnote{Remember that for $n+1 < \beta < n+2$ 
one has $c (-1)^n < 0$. This insures that the terms with
$2m > \beta -1$ are positive.} 
\begin{equation}
Z_{\mbox{\rm s}} \simeq cr {{\Delta^{\beta -1}} \over 
{[-rF'(r)]^{\beta}}} [ 1 + \sum_{m=1}^{\infty} 
{{\Gamma(1- \beta + 2m)}
\over {\Gamma(1- \beta) \Gamma(m+1) 2^m}} t^{-m}]
\label{eq6}
\end{equation}
\noindent
From the above equation one reads the susceptibility exponent
\begin{equation} 
\gamma_L = 2 - \beta + 2L
\label{eq7}
\end{equation}
\noindent
{\bf The marginal case}
\par
One again has the expansion (\ref{eq1}), 
but with $F'(r)$ set to zero.
For $\beta > 3$ the problem reduces to the generic one: it is
sufficient to replace in all formulae $\bar{\phi}$ by $r$. When
$2 < \beta < 3$ the scaling limit is defined as follows:
\begin{equation}
\Delta \to 0 \; , \; 
\Lambda \to \infty \; , \; 
t = (rc)^{{-1} \over {\beta -1}} \Lambda \mbox{\Large(} \Delta
\mbox{\Large)}^{{\beta \over {\beta-1}}}
 = \mbox{\rm const} 
\label{limit3}
\end{equation}
\noindent
In this limit the problem reduces to the solution of the equation
\begin{equation}
{{\Delta} \over {rc}}
 = (f + {{\partial} \over
 {\Lambda \partial \Delta}})^{\beta-1} \cdot 1  
\label{eq8}
\end{equation}
\noindent
with $f = - Z_s = (\Delta/rc)^{{1 \over {\beta -1}}} \chi(t)$. 
We have not succeeded in calculating in a closed form
the asymptotic expansion of $\chi(t)$. What we can offer is a
systematic recursive scheme, more efficient 
than eq. (\ref{ser}). Before
entering into further algebra and 
anticipating on the result to be obtained,
let us mention that the expansion 
is in inverse powers of $t$, which 
implies that the susceptibility exponent is
\begin{equation}
\gamma_L = {{\beta - 2} \over {\beta - 1}} 
+ {{\beta L} \over {\beta - 1}}
\label{exp3} 
\end{equation}
\par
Introduce an auxiliary variable 
$x = t^{{\beta - 1} \over \beta}$ and
rewrite (\ref{eq8}) as
\begin{equation}
x  =  {1 \over {\Gamma(1-\beta)}} 
\int_0^{\infty} ds \; s^{-\beta} e^{G(s,x)}
\label{eq9}
\end{equation}
\noindent
where $G$ is given by (\ref{gsol}) and 
$h(x) = t^{1 \over \beta} \chi(t)$. Define $u = t^{-1}$ and 
perform the substitution
\begin{eqnarray}
s & = &  \sigma u^{1 \over  \beta} \\
x & = &   u^{{1 \over  \beta}-1}
\label{subst}
\end{eqnarray}
\noindent
After simple algebra one obtains a differential equation with
analytic coefficients
\begin{equation}
[(1 - {1 \over \beta}) + {{\sigma u} 
\over \beta}] \; {{\partial G}
\over {\partial \sigma}} - u^2 \; {{\partial G}
\over {\partial u}} = - \;  (1 - {1 \over \beta}) \; \chi
\label{eq10}
\end{equation}
\noindent
We write $G$ and $\chi$ as power series in $u$ and
insert these expansions into eq. (\ref{eq10}) to get a
hierarchy of differential equations 
for the coefficients $G_n(\sigma)$.
The solution to these equations is 
$G_0(\sigma) = -  \sigma$ (since $\chi_0 = 1$) and 
\begin{equation}
G_n(\sigma) = -  \sum_{k=1}^{n+1}
{{\Gamma\mbox{\large(}{{n-k} \over 
{1 - {1 \over \beta}}} + k\mbox{\large)}} \over
{\Gamma\mbox{\large(}k + 1\mbox{\large)} 
\Gamma\mbox{\large(}{{n-k} \over 
{1 - {1 \over \beta}}} + 1\mbox{\large)}}} \;
 \sigma^k \; \chi_{n+1-k} \; , \; n \ge 1
\label{eq11}
\end{equation}
\noindent
Notice that $G_n$ depends on $\chi_k$ with $k \le n$ only. 
Finally, eq. (\ref{eq9}) becomes
\begin{equation}
1 = {1 \over {\Gamma(1-\beta)}} 
\int_0^{\infty} d\sigma \sigma^{-\beta} 
e^{- \sigma + \sum_{n=1}^\infty G_n(\sigma) t^{-n}} 
\label{eq12}
\end{equation}
\noindent
Equating to zero the coefficient of $t^{-n}$ on the right-hand side
yields $\chi_n$ in terms of $\{~\chi_k~:~k~\le~n~\}$. 

\section{Discussion}
Let us summarize what has been achieved in this work. We have
generalized the discussion of ref. \cite{bb1}, showing that BP
models fall into one of three categories. Furthermore, we have extended
the discussion to arbitrary topology. 
For each of the categories in question
we have defined an appropriate double scaling limit and written
the singular part of the partition function as a universal
asymptotic series, each term of the series corresponding to
a given topology. In the generic case, we have derived a BP equation,
valid also in the non-perturbative regime. In short, we have done 
in the context of BP models what has been earlier achieved for 
matrix models. 
\par
As in the latter case, one can relax the con\-straint $p_k \ge 0$
and con\-si\-der mul\-ti-cri\-ti\-cal model, where there exists a point
$\phi=\phi_0$ such that $U''(\phi_0) = ... = U^{[m-1]}(\phi_0) = 0$.
One obtains an equation analogous to (\ref{eq8}). However, 
now one has the integer $m$ instead of 
$\beta -1$, so that the equation is a genuine differential equation
of order $m-1$. Using the techniques developed in the last section
one defines a universal expansion of the partition function
\begin{equation}
Z_s \sim - \Delta^{1 \over m}(1 - \sum_{n=1}^\infty  \chi_n t^{-n})
 \; , \; t \sim N \Delta^{{m-1} \over m}
\label{multi}
\end{equation}
\noindent
However, the sign of the coefficients $\chi_n$ is not 
positive definite, as expected.
\par
Our motivation for studying this problem
originated from our involvement in the study of simplicial gravity.
Therefore, we would like to end this paper with remarks concerning
the hypothetical relevance of the study of branched polymers for
quantum gravity.
\par
One of the striking results obtained from computer simulations
of random geometries is that a random manifold does not stay smooth
during the simulation. It develops a tree structure: the nodes are the
baby universes and the links are the bottlenecks (wormholes) that
connect them. We shall call this tree the skeleton tree of the
manifold. For example, in 4d one observes a transition between 
two phases. In one of them the skeleton trees resemble generic BP. 
In the other the manifolds are crumpled: 
there is one big mother universe and a large number of small 
babies attached to it. Similar phase transitions are 
encountered in lower dimensions.
\par
It has been suggested in ref. \cite{bbpt} that these phase
transitions reflect the dynamics of the skeleton trees. 
Indeed, in BP theory the position of the minimum of 
$F(\phi)$ changes when one moves in the coupling space and a phase
transition occurs when this minimum hits the boundary of the support 
of this function: the generic BP turn into
crumpled structures i.e. with large or infinite Hausdorff
dimension. In sect. 2 the Hausdorff dimension has been calculated at
the tree level, but the result 
seems to hold for any fixed topology. 
\par
The hypothetical relation between the physics of 4d manifolds
and that of the associated skeletons, if confirmed by
further studies, may turn out to be a fruitful idea. The
present numerical studies of 4d random manifolds are limited 
to simplest topologies. Thus the models under study are non-unitary:
the wormholes that have been emitted cannot be reabsorbed.
These simulations, although limited to skeleton trees without loops,
are however sufficient for the 
determination of the couplings at the nodes
(the $p_k$' s). A glimpse at the unitarized theory is provided by
an extension to trees with loops, along the lines of this paper.
\par
Notice also, that a phase transition analogous to that observed
in 4d occurs in 2d, but only for large enough $c$ (or at least for
$c > 1$). But a sensible 2d gravity theory exists for $c \le 1$
only\footnote{This is at least true for generic random surface
models. Exceptions to this rule have been claimed \cite{gerv}.}.
Thus one is led to speculate that a sensible gravity theory in 4d
will only be obtained when one does in 4d something 
analogous to the reduction of the central charge in 2d.
One can develop a heuristic
argument analogous to that proposed by Cates \cite{cat} to explain
the $c=1$ barrier in 2d. Consider the continuum theory with
Einstein-Hilbert action and write the metric in the form
$g_{ab} = e^{2\sigma} \hat{g}_{ab}$. 
Assume that the BP phase in 4d is
dominated by the dynamics of the conformal factor and use the
effective action calculated in \cite{amm} 
to estimate the free energy of a
diluted gas of "spikes" : $e^{2\sigma} = 1 
+ \rho^2/ [(x - x_0)^2 + a^2] $, with $a 
\ll \rho < \mbox{\rm const} \times a$. 
As discussed in \cite{cat} the
parameter $a$ can be taken arbitrarily small, while keeping
the invariant cut-off fixed.
The free energy is
\begin{equation}
F \simeq \mbox{\large[}{{1411 + N_S + 11 N_F 
+ 62 N_V - 28} \over {360}} - 
4\mbox{\large]} \ln{{1 \over a}} \; , 
\label{fre}
\end{equation}
\noindent
where $1411$ is the contribution of transverse gravitons, 
$N_{S,F,V}$ is the number of scalar, fermion and vector 
fields respectively and $-28$ is the quantum $\sigma$ and
ghost contribution. Notice that, as emphasized in 
\cite{amm}, the sign of matter and ghost contributions 
is in 4d opposite to that found in 2d.  For pure gravity the
coefficient in front of $\ln{{1 \over a}}$ is negative and
one expects a condensation of "spikes". However the addition
of matter fields can easily change the sign of this 
coefficient and stabilize the theory. 
\vspace{0.8cm}
\par\noindent
{\bf Acknowledgements}: We wish to thank I. Antoniadis, 
K. Chadan, P.O. Mazur and E. Mottola for enlightening
conversations. One of us (J.J.) is indebted to
LPTHE for hospitality. This work was partially supported by
KBN grant 2P03B 196 02.

\end{document}